\documentclass[prl,twocolumn,aps,amssymb,superscriptaddress]{revtex4}
\usepackage{epsfig}
\usepackage{subfigure}
\usepackage{bbm}
\begin{document}

\title{Escaping from nonhyperbolic chaotic attractors}
\def\active{0}
\author{Suso Kraut} 
\affiliation{Instituto de F\'isica, Universidade de S\~ao Paulo, Caixa
Postal 66318, 05315-970 S\~ao Paulo, Brazil}

\author{Celso Grebogi}
\affiliation{Instituto de F\'isica, Universidade de S\~ao Paulo, Caixa
Postal 66318, 05315-970 S\~ao Paulo, Brazil}
\affiliation{Max-Planck-Institut f\"ur Physik komplexer Systeme, N\"othnitzer
Strasse 38, 01187 Dresden, Germany}
\begin{abstract}

The noise-induced escape process from a nonhyperbolic chaotic attractor is of 
physical and fundamental importance. We address this problem by uncovering the 
general mechanism of escape in the relevant low noise limit using the 
Hamiltonian theory of large fluctuations and by establishing the crucial role 
of the primary homoclinic tangency closest to the basin boundary in the 
dynamical process. In order to demonstrate that, we provide an unambiguous 
solution of the variational equations from the Hamiltonian theory. Our results
are substantiated with the help of physical and dynamical paradigms, such as 
the H{\'e}non and the Ikeda maps. It is further pointed out that our findings 
should be valid for driven flow systems and for experimental data. 
\end{abstract}

\maketitle
\begin{figure}[thb]
\begin{center}
\epsfig{file=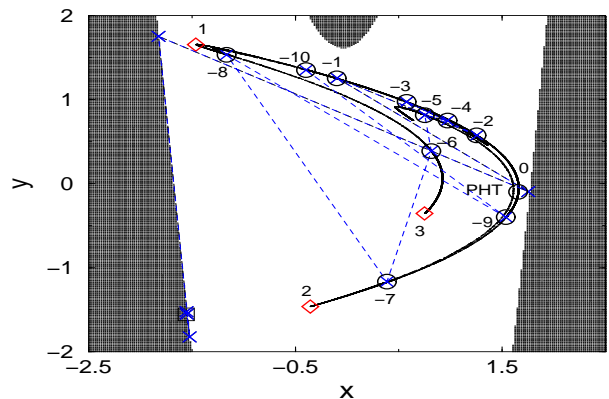,width=8cm,height=5.5cm}
\caption{(color online). CA for the H{\'e}non map with  $a=1.3$ and $b=0.29$. 
The hatched region markes the basin at infinity. The PHT at $(x,y) = 
(1.653,-0.099)$ is shown, together with its $10$-fold preimages (numbered 
-1,..,-10) as circles, the $3$-fold images of the PHT (numbered 1,2,3) as
diamonds, while the MPEP is depicted with crosses, connected through a dashed
line to guide the eyes. The saddle point on the boundary is marked with a
square.}  
\label{Henon}
\end{center}
\end{figure}
\begin{figure}[htb]
\begin{center}
\epsfig{file=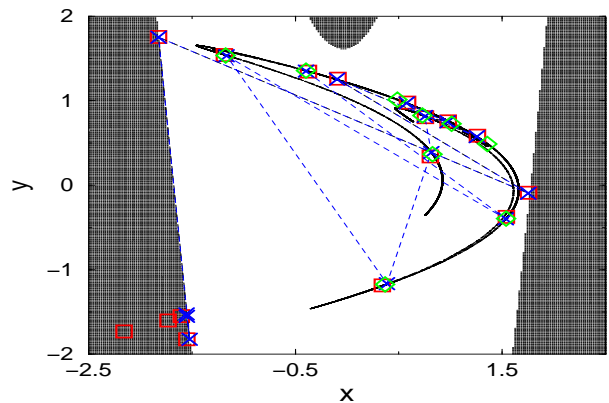,width=8cm,height=5.5cm}
\caption{(color online). The MPEP is depicted with crosses, connected through 
a dashed line to guide the eyes. The escape path, as obtained from a Monte
Carlo simulation with a noise strength of $D = 2.4 \times 10^{-4}$, is
shown with squares. An unstable period-9 orbit is displayed with diamonds.}
\label{Henon_monte_upo}
\end{center}
\end{figure}
\begin{figure}[htb]
\begin{center}
\epsfig{file=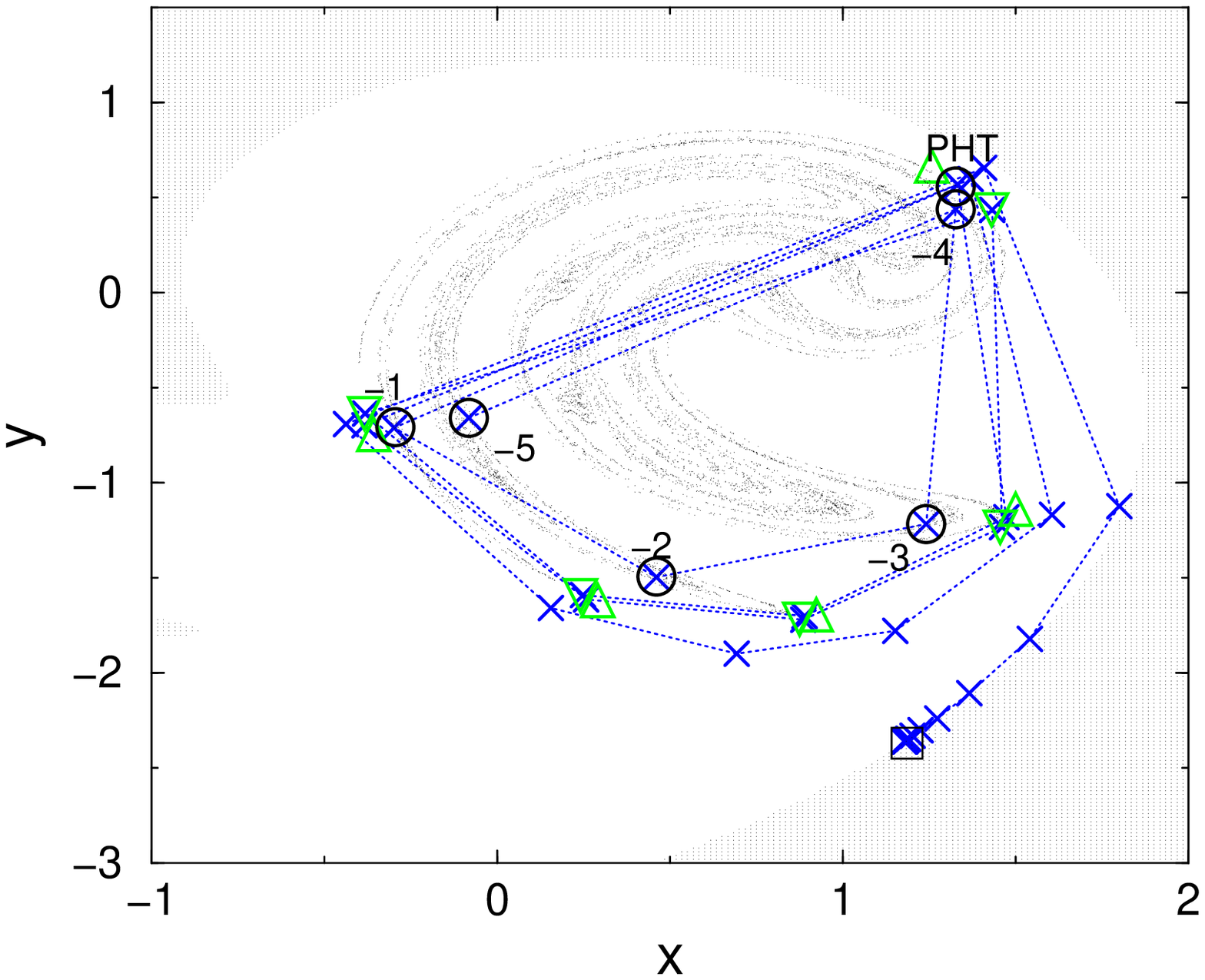,width=8cm,height=5.5cm}
\caption{(color online). CA for the Ikeda map with $a=0.92, b=0.9, \kappa=0.4$, 
and $\eta=6.0$. The hatched region marks the basin of the fixed point located 
at $(x,y)=(2.995,3.947)$. Circles depict the PHT at $(x,y) = (1.327,0.559)$ 
together with its $5$-fold preimages (numbered -1,..,-5). The MPEP is shown 
with crosses, connected through a dashed line to guide the eyes. In addition, 
two unstable period-$5$ orbits are displayed with upward and downward triangles,
respectively. The saddle point on the boundary is marked with a square.}    
\label{Ikeda}
\end{center}
\end{figure}

Many nonequilibrium systems in physics, chemistry, biology or technology 
exhibit, as a crucial feature, noise-induced escape from a metastable state. 
Examples range from Josephson junctions \cite{Devoret:1987}, switching in 
lasers \cite{Hales:2000}, Penning traps \cite{Lapidus:1999}, over chemical 
reactions \cite{Gillespie:1977} and protein folding \cite{Sali:1994} to 
electronic circuits \cite{Mantegna:1996,Luchinsky:1997}. For small noise 
intensities, a WKB-like extension of Kramers' equilibrium theory 
\cite{Kramers:1940,Hanggi:1990} has been developed to treat the realm of 
nonequilibrium systems \cite{Onsager:1953,Freidlin:1984}. This approach, 
making use of an auxiliary Hamiltonian system, identifies the so-called 
{\it most probable exit path} (MPEP), whose probability of occurrence is 
exponentially larger than that of all the other paths. The MPEP can be 
observed through its `prehistory probability distribution' and it was 
carried out numerically \cite{Dykman:1992} and experimentially 
\cite{Luchinsky:1997}. 
\\
\indent
The variational equations for the calculation of the MPEP are well known both 
for continous \cite{Freidlin:1984,Dykman:1990} and discrete systems
\cite{Kautz:1987,Beale:1989,Grassberger:1989,Graham:1991}, yet it is in general 
not clear how to solve them. Methods have been proposed only for the special 
case of escaping from periodic states (fixed points, periodic orbits or limit 
cycles) \cite{Luchinsky:2002,Beri:2003a}. Thus the challenging question of 
noise-induced escape from a {\it chaotic attractor} (CA), although tried 
before, remained to be answered. The reason is that all previous work dealing 
with the escape from a CA relies on Monte Carlo simulations of the escape path.
By making use of the `prehistory probability distribution', an unstable
periodic orbit embedded in the CA was identified and taken as initial
condition for the calculation of the MPEP
\cite{Luchinsky:1999,Khovanov:2000,Silchenko:2003}. 
\\
\indent
In this Letter we show that, in fact, the initial condition for the MPEP,
and thus the path itself, is uniquely determined by the {\it primary homoclinic 
tangency} (PHT) closest to the basin boundary, as well as its preimages and
images. Our solution of this long-standing problem does not only uncover the 
general mechanism of noise-induced escape from a nonhyperbolic CA, but it 
remedies, in addition, shortcomings in the previous method. Herewith, we 
establish the following. First, no unstable periodic orbits have to be 
determined and no Monte Carlo simulations have to be run to identify these 
orbits in the escape path. Second, the arbitrariness of the selection 
of one particular periodic orbit, whose coincidence with the MPEP is not given 
from first principles, can be dispensed with. As a matter of fact, the PHT is, 
in contradistinction, a completetly deterministic quantity and can be both 
easily calculated numerically \cite{Jaeger:1997} and extracted from experiments 
\cite{Diestelhorst:1999}. Third, complementary to the work of Ref. 
\cite{Silchenko:2003} in which the importance of the homoclinic structure of 
the fractal basin boundary to the MPEP is stressed, the present work closes 
an important gap in pointing out the essential role of the homoclinic
structure of the {\it chaotic attractor}.      
\\
\indent
Virtually all CAs occurring in nature or serving as prototype models in nonlinear
dynamics are nonhyperbolic. Nonhyperbolicity means that the stable and unstable 
manifolds of the system are tangent in the phase space. If both manifolds belong 
to the same periodic orbit, these tangencies are homoclinic. They are called 
primary, if the, generally quadratic, curvature of the manifolds in the vicinity 
of the tangent points attain a minimum. Bounded noise on nonhyperbolic CAs causes
attractor deformations \cite{Jaeger:1997,Diestelhorst:1999}, which are most
pronounced at the forward and backward iterations of the PHT closest to the
basin boundary. As a consequence, PHTs can be better defined as being those
tangencies which exhibit an amplification of a pertubation under forward and
backward dynamics.   
\\
\indent
In the following we consider discrete systems, but our results are not 
limited to them and apply equally well to any flow that can be reduced
to a Poincar{\'e} map. For the nonequilibrium case, in analogy to Kramers' 
law, the mean first exit time is given through the least action $S$
\cite{Freidlin:1984,Graham:1984,Graham:1991} by $\langle\tau \rangle \sim \exp 
\left[ \frac{S}{D}\right]$, where $D$ is the variance of the additive Gaussian
 white noise. To be specific, for a d-dimensional map ${\bf x_{n+1} = f(x_n) + 
\xi_n}$ the action of the escape trajectory to be minimized has the form
\begin{equation}
\label{action}
S_N = \frac{1}{2} \sum_{n=1}^{N}{\bf [x_{n+1}}-{\bf
f(x_n)}]^2 = \frac{1}{2} \sum_{n=1}^{N} {\bf \xi_n^T \xi_n},
\end{equation}
\\
where the ${\bf \xi_n}$ are the noise vector terms. The boundary conditions are 
such that ${\bf x_1}$ is a point of the attractor and ${\bf x_N}$ is on the
basin boundary, from where no fluctuations are needed to pass to another stable 
state. 
The MPEP, which minimizes this action, can be calculated through the Lagrangian 
\cite{Grassberger:1989,Dykman:1990}  
\begin{equation}
\label{Lagrangian}
L = \frac{1}{2} \sum_{n=1}^{N} {\bf \xi_n^T}\, {\bf \xi_n} + 
\sum_{n=1}^{N}{\bf \lambda_n^T (x_{n+1} - f(x_n) - \xi_n)}
\end{equation}
to yield upon variation of ${\bf \xi_n}, {\bf \lambda_n}$, and ${\bf x_n}$ 
the area-preserving equations
\begin{eqnarray}
\label{var1}
{\bf x_{n+1}} & = & {\bf f(x_n) + \lambda_n} \\
\label{var2}
{\bf \lambda_{n+1}} & = & {\bf \{  (Df ({x}_{n+1}))^{T}  \}}^{-1}  
\,\, {\bf \lambda_n},
\end{eqnarray}
where ${\bf Df}$ is the Jacobian matrix of ${\bf f}$. The Lagrange multipliers
${\bf \lambda_n}$ replace the noise terms ${\bf \xi_n}$. The optimal solution 
of Eqs. (\ref{var1},\ref{var2}) yields the least action 
$S = \frac{1}{2} \sum_{n=1}^{N}{\bf \lambda_n^T} \,{\bf \lambda_n}$, the 
corresponding MPEP (given by the ${\bf x_n}$) and the optimal force 
(${\bf \lambda_n}$). 
\\
\indent
The solution of these equations is intricate though, caused by wild, fractal 
fluctuations of the energy landscape having many local minima and maxima 
\cite{Graham:1984}. A way to deal with these difficulties, when escaping from 
a {\it periodic solution}, is to employ a refined shooting method 
\cite{Luchinsky:2002,Beri:2003a}. It consists of a parameterization of the
initial conditions ${\bf x_1}$ in Eq. (\ref{var1}) on a small circle 
centered in one of the components of the periodic orbit ${\bf x}_{PO}$, and 
for ${\bf \lambda_1}$ pointing to the unstable manifold of ${\bf x}_{PO}$. 
Equivalently, one can take as an initial condition one point of the periodic 
orbit itself ${\bf x_1} = {\bf x}_{PO}$ and ${\bf \lambda_1} \in r \times 
\phi$, where $\phi \in [0,2\, \pi]$ and $r$ is varied within $ r \in [r_{min},
L \, r_{min}]$. Here $L$ is the largest eigenvalue at $x_{PO}$ and $r_{min}$ 
arbitrary, but small. The upper limit of $r$ guarantees that every solution is 
only considered once. This provides an efficient numerical method for the
calculation of the MPEP.  
\\
\indent
Now, we show that the same procedure can be applied to calculate the escape from 
a {\it chaotic attractor}. To do this, one has to select the initial condition 
${\bf x_1}$ as a preimage of the PHT closest to the basin boundary, since there 
noise causes the largest deviation \cite{Jaeger:1997}. This means that the 
energy to leave the CA is the lowest. The parameterization for ${\bf \lambda_1}$
remains unchanged, since every point on the CA has a well defined largest 
eigenvalue. We first demonstrate this for one of the funamental dynamical
paradigms, the H{\'e}non map \cite{Henon:1976} 
\begin{eqnarray}
x_{n+1} & = & a - x_n ^2 + b\,y_n + \xi_n\\ 
y_{n+1} & = & x_n. \nonumber
\end{eqnarray}
We choose the parameters $a=1.3$ and $b=0.29$ for which a CA and another 
attracting state at infinity coexist. In Fig. \ref{Henon} the CA is plotted, 
as well as its basin of attraction, and the saddle point on the boundary 
(square). We also include the PHT closest to the boundary together with its 
$10$-fold preimages (circles) and the MPEP (crosses). The latter is found by 
iterating Eq. (\ref{var1}) at the $10$-fold preimage of the PHT and 
by looking for the absolute minimum when changing ${\bf \lambda_1}$ as 
described above. As it can be seen, the MPEP moves initially very close to the
deterministic dynamics yet it deviates increasingly with every iteration (as 
expected, since otherwise there were no escape).     
At the PHT the MPEP already differs considerably from it. Only there, it is 
for the first time located {\it outside}  the CA and advances then, following
the elongated images of the PHT (depicted as diamonds), to the basin boundary. 
From there it approaches the saddle point located on the basin boundary. We
emphasize that it is not presupposed that the MPEP leaves the CA in the vicinity
of the PHT. We simply use a preimage of the PHT as the initial condition 
${\bf x_1}$ and look for the optimal solution of Eqs. (\ref{var1},\ref{var2}). 
The fact that the MPEP is indeed following very closely the structure of the PHT
(preimages and images) is thus a confirmation of our claim. 
\begin{figure}[h]
\begin{center}
\epsfig{file=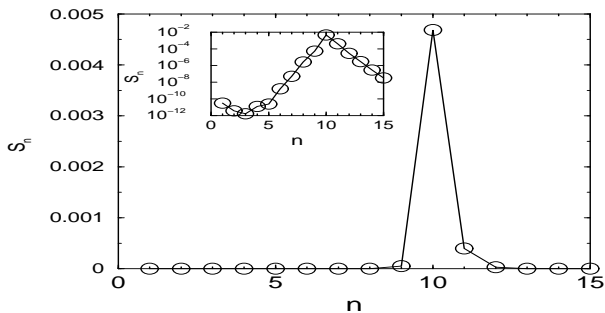,width=8cm,height=4.25cm}
\caption{The stepwise action $S_n=\frac{1}{2}{\bf\lambda_n^T}\,{\bf\lambda_n}$
for the H{\'e}non map. At each time step, the value of the graph corresponds to
the optimal force ${\bf\lambda_n}$, as a result of the optimal solution of 
Eqs. (\ref{var1},\ref{var2}). The inset shows the data in semi-log scale.} 
\label{Energy_Henon}
\end{center}
\end{figure}
\\
\indent
To shed further light on this general mechanism, the stepwise action 
$S_n=\frac{1}{2}{\bf \lambda_n^T}\,{\bf \lambda_n}$ is plotted in Fig.
\ref{Energy_Henon}. There is only one large peak (step $10$) where the MPEP 
moves away from the CA close to the PHT. Afterwards, the required energy 
drops off exponentially, since the MPEP follows a relaxational path. It is 
important to note that this scenario differs from the one of the attractor 
deformation. In the latter, the trajectory deviates from the CA {\it starting 
at the PHT}, while in our case the MPEP departs already at the preimages
(cf. Fig. \ref{Energy_Henon}, inset) and is, when passing the PHT, a {\it
finite distance} away. This is the reason why we have to start iterating
Eq. (\ref{var1}) at some preimage of the PHT. However, also by using only the 
$5$-fold preimage of the PHT the same MPEP was obtained. Thus neither the 
exact number of preimages nor the accurate determination of the PHT is 
crucial - an indication about  the robustness of the method.  
\\
\indent 
In Fig. \ref{Henon_monte_upo} the MPEP is compared with a direct Monte Carlo 
simulation of the escape trajectory. The simulation was carried out with a 
noise strength of $D  = 2.4 \times 10^{-4}$, resulting in a trajectory of 
length $2.5 \times 10^{12}$. One sees a strikingly good agreement of about $15$ 
iterations before the path approaches the saddle point, reaching back to the
$10$-th preimage of the PHT. This remarkable coincidence of the simulated path
with the MPEP corroborates the general validity of the procedure.
\\
\indent 
To connect the MPEP with the low-period unstable periodic orbits embedded in 
the CA, we calculate all periodic orbits up to period $20$ \cite{Biham:1989}. 
Only one periodic orbit is found to be close to the MPEP, having period $9$. 
It is also included in Fig. \ref{Henon_monte_upo}. Thus, if one had adopted 
the previous method of finding periodic orbits close to the MPEP by using
stochastic simulations of the system, one might have been led to conclude that 
the MPEP starts at that period $9$ orbit. However, such a reasoning obscures 
the general deterministic structure of the MPEP with respect to the PHT, as
presented in this work. 
\\
\indent
Next we exemplify the method with the Ikeda map \cite{Ikeda:1979}, a
discrete model of a laser pulse in an optical cavity. With complex
variables it has the form 
\begin{equation}
{\bf z_{n+1}} = a + b {\bf z_n} \exp \left[ i \kappa - \frac{i \eta}{1 + |{\bf
z_n}|^2} \right] + {\bf \xi_n},
\label{Ikeda_map}
\end{equation}
where ${\bf z_n} = x_n + i y_n$ is related to the amplitude and phase of the 
$nth$ laser pulse exiting the cavity. The parameter $a$ is the laser input
amplitude and ($1-b$) the damping, while the empty cavity detuning is given by
$\kappa$ and the detuning due to a nonlinear dielectric medium by $\eta$.
\\
\indent
We fix the parameters at $a = 0.92, b = 0.9, \kappa = 0.4$ and $\eta =6.0$. 
For the noisefree system, two stable states are present, a fixed point and a
CA. In Fig. \ref{Ikeda} we present the result for the Ikeda map, analogously 
to Fig. \ref{Henon}. The CA, the basin of attraction and the saddle point on 
the boundary (square) are depicted. It is also shown the PHT closest to the 
boundary together with its $5$-fold preimages (circles) and the MPEP (crosses). 
We include two period-5 orbits as well (triangles), which are located near 
the images of the PHT. The behaviour of the MPEP and the optimal force 
(not shown) is qualitatively the same up to where the path reaches 
the PHT. Then it passes very close to two unstable period-5 orbits, which are 
{\it outside} the CA, before it approaches the basin boundary. This is so 
because they happen to be located exactly at the regions of phase space where 
the noise elongations attain a maximum, which occurs close to the 
images of the PHT \cite{Jaeger:1997}. The passing of the MPEP through another 
invariant set was also found in a different system \cite{Khovanov:2000}, as 
well in the more complex situation when there exists a chaotic saddle 
\cite{Kraut:2003a}. It does not happen for the H{\'e}non map though, since all 
the unstable periodic orbits lie, for the present parameter values, on the CA.  
\\
\indent
To conclude, we have unveiled the general mechanism of noise-induced escape
from a nonhyperbolic CA. It is shown that the MPEP exits at the vicinity of
the PHT closest to the basin boundary. 
Our findings are established by solving unambiguously the auxiliary 
Hamiltonian system, which yields the exact description in the low noise limit. 
This has, for the first time, been established without taking recourse to 
Monte Carlo simulations, and only the knowledge of the deterministic structure 
of the PHT is required. This mechanism gives a robust practical procedure and 
should also be able to be confirmed experimentally 
\cite{Luchinsky:1997,Diestelhorst:1999}.  
It is advantageous both for stabilizing systems and for energy-optimal
switching between different states (cf. \cite{Khovanov:2000}). 
\\
\indent
We acknowledge D. G. Luchinsky, S. Beri, U. Feudel, M. S. Baptista, and H. Kantz 
for valuable  discussions. This work was supported by the Alexander von Humboldt
Stiftung. 


\end{document}